# Thermoelectric power factor under strain-induced band-alignment in the half-Heuslers NbCoSn and TiCoSb

Chathurangi Kumarasinghe[*], and Neophytos Neophytou

School of Engineering, University of Warwick, Coventry, CV4 7AL, UK

[*] Chathu.Kumarasinghe@warwick.ac.uk

## Abstract

Band convergence is an effective strategy to improve the thermoelectric performance of complex bandstructure thermoelectric materials. Half-Heuslers are good candidates for band convergence studies because they have multiple bands near the valence bad edge that can be converged through various band engineering approaches providing power factor improvement opportunities. Theoretical calculations to identify the outcome of band convergence employ various approximations for the carrier scattering relaxation times (the most common being the constant relaxation time approximation) due to the high computational complexity involved in extracting them accurately. Here, we compare the outcome of strain-induced band convergence under two such scattering scenarios: i) the most commonly used constant relaxation time approximation and ii) energy dependent inter- and intra-valley scattering considerations for the half-Heuslers NbCoSn and TiCoSb. We show that the outcome of band convergence on the power factor depends on the carrier scattering assumptions, as well as the temperature. For both materials examined, band convergence improves the power factor. For NbCoSn, however, band convergence becomes more beneficial as temperature increases, under both scattering relaxation time assumptions. In the case of TiCoSb, on the other hand, constant relaxation time considerations also indicate that the relative power factor improvement increases with temperature, but under the energy dependent scattering time considerations, the relative improvement weakens with temperature. This indicates that the scattering details need to be accurately considered in band convergence studies to predict more accurate trends.

## Introduction

Thermoelectric materials are capable of direct conversion of thermal energy into electricity and vice versa. They have drawn interest as an environment-friendly and highly reliable energy conversion technology[1–4]. The efficiency of a thermoelectric material is quantified by the thermoelectric figure of merit (ZT) given by:

$$ZT = (\sigma S^2 T)/\kappa \; , \qquad\qquad (1)$$

where $\sigma$ is the electrical conductivity, S is the Seebeck coefficient, T is the temperature and $\kappa$ is the thermal conductivity of the material. To obtain a high ZT, a high electrical conductivity, a high Seebeck coefficient, i.e. a high power factor ($\sigma S^2$), and a low thermal conductivity $\kappa$ is needed, however, simultaneous optimization of these parameters is difficult due to their complex inter-dependencies.

Bandstructure engineering approaches such as 'band-convergence' also known as 'band-alignment', can overcome these inter-dependencies and improve the power factor (PF)[5,6]. The aim of band-convergence is to increase the number of carriers contributing to electronic transport by increasing the valley or orbital degeneracy near conduction or valence bands edges. In bulk materials, bandstructures can be manipulated to achieve this by applying strain, doping, alloying, and second phasing with other suitable structures [7–9]. For example, half-Heusler (HH) alloys [10–14] have complex bandstructures with the potential for beneficial band convergence. They are also known to have good thermal and mechanical stability, large scale reproducibility, and made out of relatively inexpensive, non-toxic elements.

One of the most commonly used theoretical approaches for obtaining thermoelectric properties of materials is applying the Boltzmann transport theory in the relaxation time approximation using detailed material bandstructures obtained from *ab initio* density functional theory (DFT) [15,16]. The carrier relaxation time is an important input in this calculation, but due to the high computational complexity in calculating accurate carrier relaxation times, a constant relaxation time is generally assumed (usually $\tau \approx 10^{-14} s$ is used). However, such simplifications are known to give less accurate predictions particularly during band-convergence optimization studies [17]. This is because, while band-alignment can increase the number of carriers available for conduction, it can also increase the number of states that carriers scatter into. Therefore, the energy dependence of the scattering mechanisms, as well as intra- or inter-valley scattering considerations are important in identifying if a given bandstructure engineering approach leads to an improved power factor. In addition, the scattering rate depends on the temperature as well with more scattering of electrons/holes by phonons expected at higher temperatures.

In this work we study the effect of considering different scattering time scenarios on the strain-induced band-convergence for two Co-based half-Heuslers, NbCoSn and TiCoSb. We consider constant ($\tau_C$), and energy dependent inter- and intra- valley scattering ($\tau_{IIV}$) relaxation time approximations, and quantify the relative power factor improvements that each predict at different temperatures and strain levels. We finally stress the importance of accurate treatment of scattering in arriving to useful predictions.

## Methods

**Boltzmann Transport Theory:** Thermoelectric coefficients are calculated using the Boltzmann transport formalism within the relaxation time approximation (RTA). We describe the material properties using bandstructures obtained from DFT calculations. The thermoelectric coefficients, electrical conductivity $\sigma_{\alpha\beta}(T, E_F)$ and Seebeck coefficient $S_{\alpha\beta}(T, E_F)$ tensors, can be written as [18,19]:

$$\sigma_{\alpha\beta}(T, E_F) = e^2 \int \Xi_{\alpha\beta}(E) \frac{\partial f(E, E_F, T)}{\partial E} dE \qquad (2)$$

$$S_{\alpha\beta}(T, E_F) = \frac{e}{T\sigma_{\alpha\beta}(T, E_F)} \int \Xi_{\alpha\beta}(E)(-E_F) \frac{\partial f(E, E_F, T)}{\partial E} dE \qquad (3)$$

where $f(E, E_F, T)$ is the fermi distribution function at a given temperature $T$ and a chemical potential level $E_F$ and $e$ is the charge of an electron. $\Xi_{\alpha\beta}(E)$ is the transport distribution (TD) function, which is given by:

$$\Xi_{\alpha\beta}(E) = \sum_{i,\mathbf{k}} \tau_{i,\mathbf{k}}(E) v_\alpha(i,\mathbf{k}) v_\beta(i,\mathbf{k}) \, \delta\big(E - E_{i,\mathbf{k}}\big) , \tag{4}$$

where $i$ and $\mathbf{k}$ represent the band index and the k-point, respectively. $\tau_{i,\mathbf{k}}(E)$ is the electron relaxation time and $v_\alpha(i,\mathbf{k})$ ($\alpha = \{x, y, z\}$) represents the $\alpha^{th}$ component of the group velocity $\mathbf{v}(i,\mathbf{k})$, which can be derived from the gradient of the bands in the bandstructure as:

$$\mathbf{v}(i,k) = \frac{1}{\hbar} \, \nabla_{\mathbf{k}} \, E_{i,\mathbf{k}} . \tag{5}$$

**Ab initio electronic structure calculations:** *Ab initio* DFT calculations were performed for the Co-based HHs, NbCoSn and TiCoSb with the QUANTUM ESPRESSO package [20]. Projector augmented wave technique was used with the PBE-GGA functional. Throughout our calculations, a kinetic energy cutoff greater than 60Ry was used for wavefunctions and an energy convergence criterion of $10^{-8}$ Ry was adopted for self-consistency. For transport property calculations, a 15x15x15 Monkhorst–Pack k-point sampling was used for the primitive unit cell with three atoms. Calculations using denser k-points were carried out to confirm the convergence of the results.

**Relaxation time approximation**: By applying the Boltzmann transport theory using the DFT extracted bandstructures, we examine the influence of the nature of scattering times on the power factor for various temperatures and upon strain-induced band-convergence. We consider: i) the commonly employed constant relaxation time ($\tau_i = \tau_C$) and ii) scattering proportional to total density of states, allowing both inter- and intra-band scattering ($\tau_i(E) = \tau_{IIV} \propto 1/\sum_i \mathrm{DOS}_i(E)$).

To obtain a general understanding about these two scattering scenarios, we first analyse them using a simple parabolic band model. Since under the parabolic band approximation, the velocity and density of states of each band is $v_i(E) =$ $(2E/m_i)^{1/2}$, and $\mathrm{DOS}_i(E) = 2^{1/2} m_i^{3/2} N_i E^{1/2}/(\pi^2 \hbar^3)$, respectively, the TD function given in Eq.4 for valence bands is reduced to:

$$\Xi(E) \propto \sum_i N_i \tau_i(E) \, m_i^{\frac{1}{2}} E^{\frac{3}{2}} \; (\mathrm{H}[-\Delta E_i]). \qquad (6)$$

where the subscript $i$ indicates each band, $\Delta E_i$ indicates the distance to the band edge from the valence band edge, $N_i$ indicates the band degeneracy and $\mathrm{H}[\Delta E_i]$ indicates the Heaviside step function. (For conduction bands, $(\mathrm{H}[-\Delta E_i])$ should be replace by $\mathrm{H}[\Delta E_i]$.)

Under the constant relaxation time approximation, $\tau_i(E) = \tau_C$, where $\tau_C$ is a constant. Therefore, the TD function relation given by Eq. (6) can be further simplified to:

$$\Xi(E) \propto \sum_i m_i^{\frac{1}{2}} E^{\frac{3}{2}} \mathrm{H}[-\Delta E_i]. \qquad (7)$$

This indicates that under a constant relaxation time, aligning a band of any mass will increase the TD function, resulting finally in an increased conductivity, with the larger masses resulting in larger improvements, a common scenario seen in most band-alignment theory literature. The magnitude and sign of the Seebeck coefficient are related to an asymmetry of the electron transport around the Fermi level [21,22], which is indicated by the energy gradient of the TD function. If the band convergence does not additionally introduce significant asymmetry (significant change in gradient of the TD function), there will not be a significant change in the Seebeck coefficient. Therefore, in the case of a constant relaxation time, a power factor improvement is always achieved upon band alignment, driven by improvements in the conductivity, with a heavier aligning band being preferred, as it offers a higher density of conducting states (but not more scattering states under $\tau_C$). As we discuss further below, this is not the case when $\tau = \tau_{IIV}(E)$.

In the case of inter- and intra-band scattering ($\tau_{IIV}(E)$) that we consider next, carriers are allowed to scatter elastically to the total density of states available at the energy under consideration, without any selection rules, i.e. both intra and inter-band (with inter and intra-valley in multi-valley materials) scattering is allowed ($\tau_i(E) = \tau_{\mathrm{IIV}}(E) \propto 1/\sum_i \mathrm{DOS}_i(E)$). The TD functions relation given by Eq. (6) in this case can be simplified to:

$$\Xi(E) \propto \frac{\sum_i m_i^{\frac{1}{2}} E^{\frac{3}{2}} \mathrm{H}[-\Delta E_i]}{\sum_i m_i^{\frac{3}{2}} E^{\frac{1}{2}} \mathrm{H}[-\Delta E_i]} \qquad (8)$$

From Eq. (8), since the denominator (its appearance being a result of the scattering times being inversely proportional to the DOS), has a higher mass exponent, it can be deduced that upon full band alignment, the TD function will only increase when a light band is brought close to the band edge and is aligned with a heavier band. When additional bands are gradually brought close to the band edge to be aligned, three competing effects take place: i) the presence of the additional conducting states tends to increase the TD function, ii) the same states increase the scattering of already aligned bands, which tends to reduce the TD function, and iii) scattering from newly aligned bands reduces since there are less states to scatter into at energies closer to valence band edge (due to the energy dependence $E^{1/2}$ of the DOS), increasing the TD function. These interdependences do not allow for significant improvements in the TD, the conductivity, and the PF as in the previous two scattering scenarios. As a result of these competing effects, aligning bands is not always advantageous for the power factor under $\tau_{\mathrm{IIV}}(E)$[17].

**Temperature dependent carrier relaxation time**: When performing temperature dependent studies, and evaluating the trends under increasing temperature, the Fermi distribution broadens, which allows the occupation of more states, however, to satisfy charge neutrality the Fermi level will shift towards the bandgap, increasing the Seebeck coefficient. If one compares at the same Fermi level position, rather than carrier density, however, the conductivity increases, while not many changes are observed in the Seebeck coefficient. On the other hand, temperature increases the population of phonons, which increase carrier-phonon scattering. In the most common acoustic deformation potential scattering scenario, under the equipartition approximation, the scattering rate increases linearly with temperature. Thus, we account for the increase in carrier scattering with increase in temperature by scaling the relaxation times by $1/T$. We refer to this as 'T-normalized relaxation time'. In the case of the constant relaxation time, we take $\tau_C = 10^{-14}$ s at 300 K, and linearly scale for other temperatures as $\tau_C = 10^{-14} \left(\frac{300}{T}\right) s$, to account for deviations from room temperature In the case of energy dependent relaxation time, temperature scaling is implemented as $\tau_{IIV} \propto 1/\sum_i \mathrm{DOS}_i(E)\,(300/T)$.

## Results and Discussion

Figure 1 shows the power factors calculated using both temperature normalized and non-normalized relaxation times. Under non-normalized conditions (solid lines), at the same Fermi level position, the power factor rises with the temperature because the derivative of the Fermi distribution function $((\partial f(E, E_F, T))/\partial E)$ in the Eq. 2 broadens (for a given Fermi level), increasing the conductivity. However, when we take into account the increase in carrier scattering by normalizing the carrier scattering time by $300/T$, the power factor difference almost disappears, and only modest differences exist between temperatures for both scattering scenarios. This illustrates that it is important to take increased carrier scattering into consideration when comparing thermoelectric parameters at different temperatures. (Note that it is often the case in experiments to observe that the power factor changes with temperature. That is because those are performed at a constant density, rather than a constant Fermi level position, which would have shifted the two lines compared to each other).

Band convergence can be achieved using a variety of methods including applying strain [17,23,24] and alloying[6,25,26]. Here, for the purposes of our investigation into the influence of band alignment on the PF under different scattering scenarios and temperatures, we use the easiest method within DFT, which is the use of hydro-static strain, either compressive or tensile.

In Fig. 2(c), in the unstrained NbCoSn bandstructure, it is apparent that multiple bands from several valleys are available close to the valence band edge $VB_0$. Bands at the L and W points are already aligned at $VB_0$. There also exist heavy and light bands at the X and $\Gamma$ points within 0.3eV of the $VB_0$. Aligning these bands that are in the vicinity of the $VB_0$, particularly the bands at the X point that have a large equivalent valley degeneracy of 3, can lead to an improved conductivity and power factor. As seen in Figure 2(a-d), the bands of NbCoSn can be manipulated with compression and expansion. When the material is compressed, the bands at the X point are brought closer to $VB_0$, reducing their energy separation $\Delta E$, and when expanded, $\Delta E$ increases. It is important to note, however, that the curvatures of the bands, i.e. the carrier effective masses also change with strain. Compression increases the band curvatures (lower carrier effective mass) in general and expansion has the opposite effect. The fact that the masses are reduced with band convergence is unfavorable under a constant scattering rate as seen above in Eq.7. It is, however, favorable under $\tau_{IIV}(E)$ under most situations[17] as it brings bands with higher velocities that scatter less within the transport window. We consider large strain values of up to 5% and achieve a degree of alignment, but a compressive strain larger than 5% is required to completely align the bands.

We now examine the thermoelectric coefficients of NbCoSn under strain, for three different temperatures 300K, 600K, and 900K (Fig. 3). Figure 3 shows the TE coefficients conductivity, Seebeck coefficient and the power factor calculated under the assumption of a constant scattering time ($\tau_C$). Later on, we will show how these observations are altered in the case of energy dependent $\tau_{IIV}$ scattering. At 300K (Fig. 3a-c) we only see a 3% improvement in the PF with band convergence. With increasing temperature, the conductivity for non-degenerate Fermi levels increases, while it decreases for higher Fermi levels. The opposite effect is seen in the Seebeck coefficient where its values decrease for non-degenerate conditions and increase for highly degenerate Fermi values (right versus left regions of the sub-figures). Also, the gradients of conductivity and Seebeck coefficient curves are reducing with increasing temperature. The combined effect is that with increasing temperature the power factor improves for all compressive strain value, which tend to align the bands, with the improvements at 5% compressive strain to increase from 3% to 13% to 21% at temperatures of 300 K, 600 K, and 900 K, respectively. Note that the absolute PF values do not necessarily increase with temperature, it is the relative increase between the unstrained and strained cases that increases.

Next, we examine the behavior of NbCoSn under the energy dependent $\tau_{IIV}$ scattering for the three temperatures of interest, 300 K, 600 K, and 900 K (Fig.4, column-wise). In this case, the differences in the conductivity are more noticeable between the different strain conditions, compared to the $\tau_C$ case in Fig. 3. The Seebeck coefficients as in the $\tau_C$ case are almost unchanged with strain at the same Fermi level position, especially for the higher temperatures. The improvements to the PF, on the other hand, are larger with compressive strain, and they begin to appear and being significant even at 300 K. In this $\tau_{IIV}$ case, the PF improvements for the 5% compressive strain, at 300 K, 600 K, and 900 K are 20%, 23%, and 33%, respectively.

The larger improvements compared to the $\tau_C$ case, originate from the changes in the curvature of the bands in addition to the reduced band separation. The bands that are brought closer to the band edge to be aligned as well as the bands already at the band edge become lighter in the process in NbCoSn [17]. Specifically, the masses of the bands at X valley band, changes from $0.48m_0$ to $0.29m_0$ while its separation is reduced from 315 meV to 232 meV upon 5% compression. Alignment tends to increase the PF in both $\tau_C$ and $\tau_{IIV}$, but in the $\tau_C$ case the fact that the bands become lighter, mitigates this benefit. On the other hand, in the case of $\tau_{IIV}$, light bands are beneficial, which provides an additional improvement to the PF. Finally, the reason the benefits increase with temperature in both cases, is due to the simple fact that the broadening of the Fermi distribution with temperature, allows for an exponentially larger percentage contribution of the upper valleys to the total contribution. The separation $\Delta E$ is still large, even at 5% compression, accounting for

several k$_B$$T$ even at 900 K. Thus, at the same Fermi level position, the overall relative PF improvement with temperature increases with $T$. For NbCoSn, either $\tau_C$ or $\tau_{IIV}$ indicate a higher benefits of band alignment at higher temperatures.

Next, we apply strain to TiCoSb in the same way. The smaller energy separation $\Delta E$ between bands at L and $\Gamma$ points (see Fig. 5c) can be reduced by applying compressive strain, leading to band convergence, as shown in Fig. 5 (a-d). The unstrained $\Delta E$ value in this situation is only 40 meV ( 1.55 $k_B T$ at $T = 300$ K), i.e. the bands are almost aligned even without strain. The bands can be fully aligned by applying only ≈2% compressive strain as shown in Fig. 5b (a much more realistic value compared to the one needed for NbCoSn).

Figure 6 shows the TE coefficients conductivity, Seebeck coefficient and PF for the constant time scattering assumption $\tau_C$ for 300 K, 600 K, and 900 K columnwise, in the same manner as earlier for the NbCoSn. We only see 6% improvement when bands are fully aligned using compressive strain (Fig. 6(c)), but the improvement increases for higher temperatures, reaching 13% and 19% at 600 K and 900K, respectively (Fig. 6(f), 6(i)). Again, the improvement in the PF with compression at a given temperature originates from the conductivity. Improvements increase in the higher temperatures because the positive contribution from the broadening of the $\partial f(E, E\_F, T))/\partial E$ function overweight the negative effects on the TD function from increased scattering.

We now examine the case where the relaxation time in TiCoSb is energy dependent ($\tau_{IIV}$). Figure 7 shows the TE coefficients conductivity, Seebeck coefficient and PF for this energy dependent scattering assumption for the three temperatures of interest, under 5% expansion, 2% compression, and 5% compression as before in Fig. 6. We observe a 37% improvement when bands are fully aligned using compressive strain (Fig. 7(a)) at 300 K. The fact that band curvatures also increase (carrier effective masses reduce) with alignment, is more favorable under the $\tau_{\text{IIV}}$ scattering scenario, which has contributed to this large improvement observed. Interestingly, as opposed to all previous scenarios the improvement reduces with temperature increase, down to 27% and 25% at 600 K and 900 K, respectively (Fig. 6(f), 6(i)). The reason behind the smaller improvement variation with temperature upon band alignment for TiCoSb compared to the NbCoSn case, is of course the fact that the bands were closer together to begin with, and thus in all temperature cases the initial and the aligning bands contribute similarly. In other words, the influence of the broadening of $\partial f(E, E\_F, T))/\partial E$ affects all cases similarly. The increase in scattering with temperature, however, overcomes any positive contribution from this broadening, reducing the PF improvement with temperature.

In summary, as seen in Fig.8, band convergence is advantageous for both NbCoSn and TiCoSb (there is an improvement in the PF). Under both scattering scenarios we have considered, band convergence provides larger relative benefits for NbCoSn as the temperature increases (blue lines). Band convergence on the other hand, provides relative improvements in the PF for TiCoSb under $\tau_{\text{C}}$, but reductions under $\tau_{\text{IIV}}$ as the temperature increases.

## Conclusion

In this work we have used strain to align the bands near the valence band edge of the p-type Co-based half-Heuslers NbCoSn and TiCoSb for thermoelectric power factor improvements. Using the Boltzmann transport equation under the relaxation time approximation, we explored the band alignment effect on the power factor under two different scattering conditions (as the detail scattering physics of half-Heuslers are still not known): i) the constant relaxation time approximation – as is common in the literature, ii) scattering rates depending on energy, proportional to the density of final states, with both inter- and intra-band\ inter- and intra-valley scattering considerations. We showed that the outcome of band alignment can be different in each of the different scattering cases and at different temperatures for these materials. It depends on the nature of the bandstructure, particularly on how the curvatures change with band convergence and the relative separations between bands to begin with. When benefits of broadening of $\partial f(E, E\_F, T))/\partial E$ overcome negative contributions from increased carrier scattering with temperature, we see increasing improvements in the PF with temperature. For NbCoSn, band convergence using strain is more beneficial in higher temperatures in both scattering scenarios, while for TiCoSb higher temperatures are only beneficial under a constant rate of scattering. Our work stresses the importance of more accurate theoretical treatment of carrier relaxation times and their temperature dependences.

### Acknowledgements

This work has received funding from the European Research Council (ERC) under the European Union's Horizon 2020 Research and Innovation Programme (Grant Agreement No. 678763).

# Figures

Figure 1

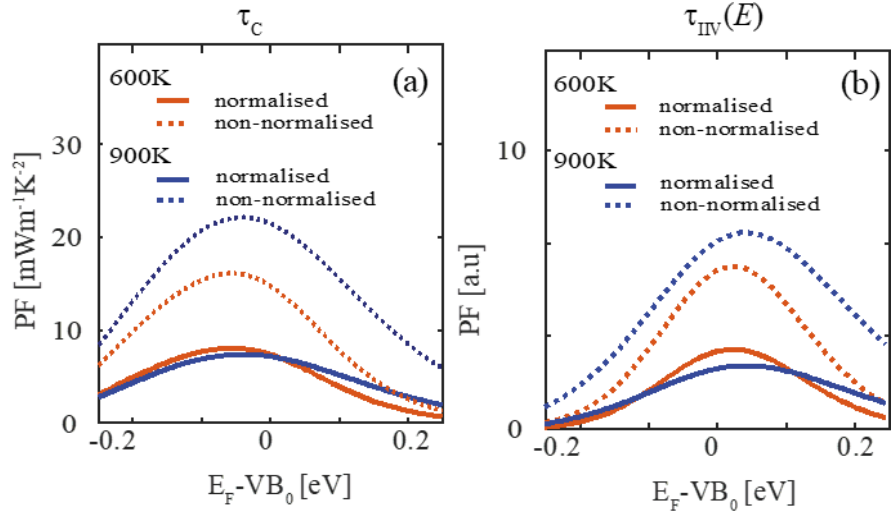

**Fig. 1. Power factors calculated using normalized (solid-lines) and non-normalized(dashed-lines) carrier relaxation times at 600K (red) and 900K(blue) for (a) constant relaxation time ($\tau_C$) and (b) energy dependent relaxation time ($\tau_{IIV}$) for NbCoSn.**

Figure 2

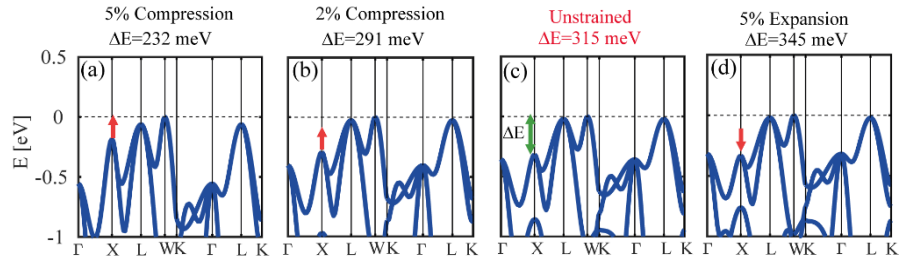

**Fig. 2. (a)-(d) Application of compressive and tensile strain in NbCoSb to align the X valley with the valence band edge. Strain values and the energy separation ΔE between the X valley and VB0 are noted above the sub-figures. Subfigure (d) shows the unstrained bandstructure.**

Figure 3

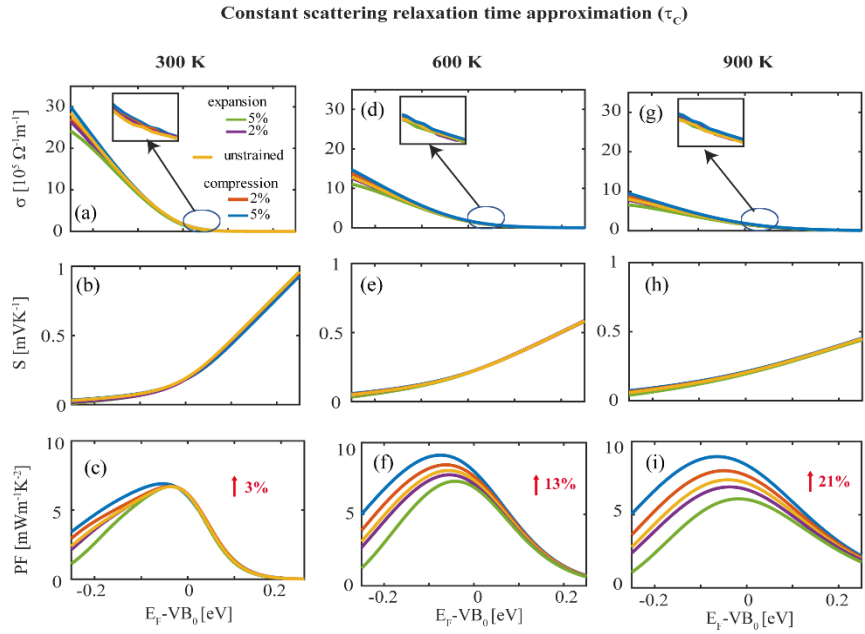

**Fig. 3. Thermoelectric coefficients (σ, S, and PF) calculated for NbCoSn using the temperature-normalized constant relaxation time approximation: (a)-(c) for 300K, (d)-(f) for 600K and (g)-(i) for 900K. The percentage improvement given is the peak to peak improvement between the unstrained (ΔE=315 meV) and 5% compressive strain (ΔE=232 meV), given by the yellow-solid lines and blue-solid lines, respectively.**

Figure 4

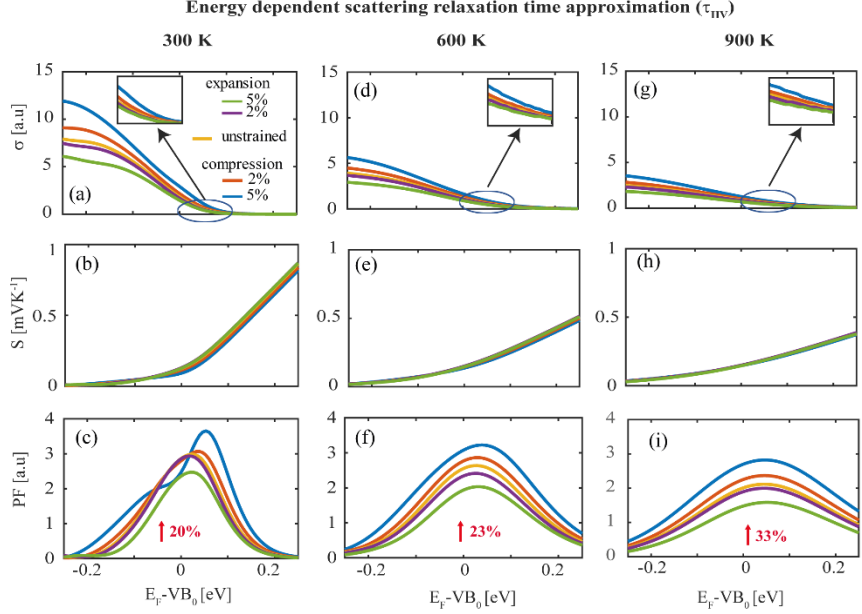

**Energy dependent scattering relaxation time approximation (τ_IIV)**

Fig. 4. Thermoelectric coefficients (σ, S, and PF) calculated for NbCoSn using the temperature-normalized energy dependent $\tau_{IIV}$ relaxation time approximation: (a)-(c) for 300K, (d)-(f) for 600K and (g)-(i) for 900K. The percentage improvement given is the peak to peak improvement between the unstrained (ΔE=315 meV) and 5% compressive strain (ΔE=232 meV), given by the yellow-solid lines and blue-solid lines, respectively.

Figure 5

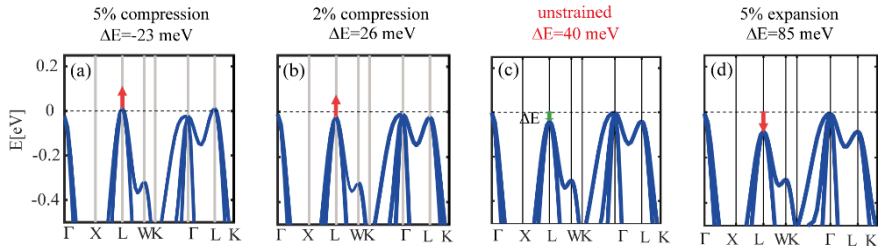

Fig. 5. (a)-(d) Application of compressive and tensile strain in TiCoSb to align the X valley with the valence band edge. Strain values and the energy separation ΔE between the X valley and VB0 are noted above the sub-figures. Subfigure (d) shows the unstrained bandstructure.

Figure 6

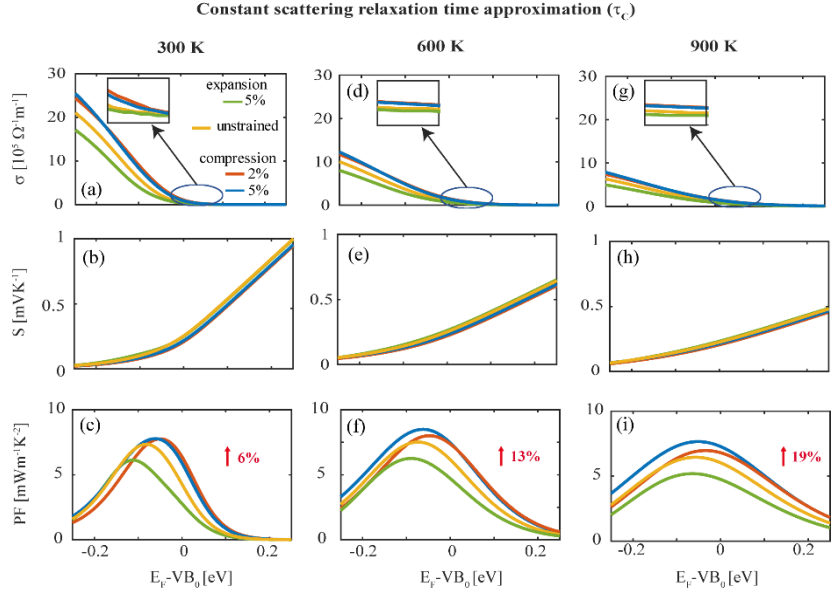

**Fig. 6. Thermoelectric coefficients (σ, S, and PF) calculated for TiCoSb using the temperature-normalized $\tau_C$ approximation: (a)-(c) for 300K, (d)-(f) for 600K and (g)-(i) for 900K. The percentage improvement given is the peak to peak improvement between the unstrained (ΔE=40 meV) and 5% compressive strain (ΔE=-23 meV), given by the yellow-solid lines and blue-solid lines, respectively.**

Figure 7

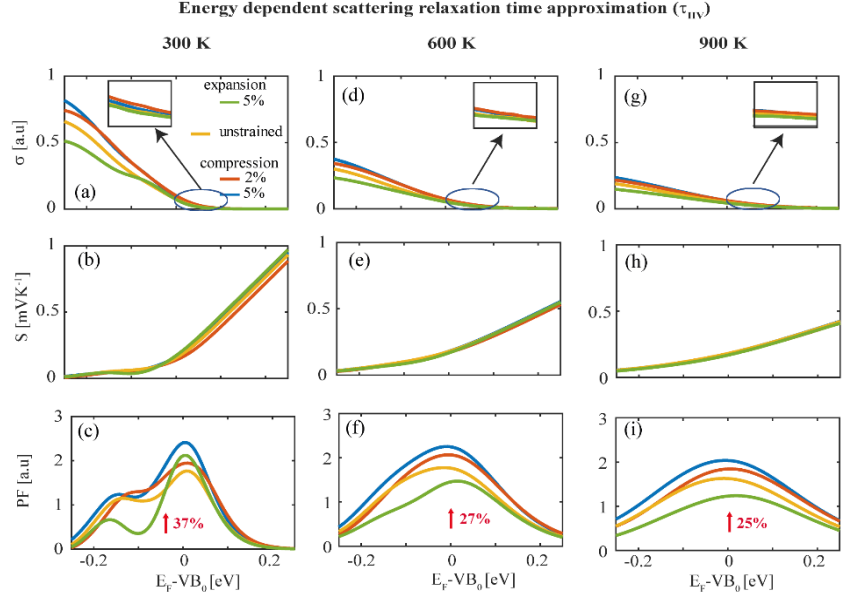

**Fig. 7. Thermoelectric coefficients (σ, S, and PF) calculated for TiCoSb using the temperature-normalized $\tau_{IIV}$ relaxation time approximation: (a)-(c) for 300K, (d)-(f) for 600K and (g)-(i) for 900K. The percentage improvement given is the peak to peak improvement between the unstrained (ΔE=40 meV) and 5% compressive strain (ΔE=-23 meV), given by the yellow-solid lines and blue-solid lines, respectively.**

Figure 8

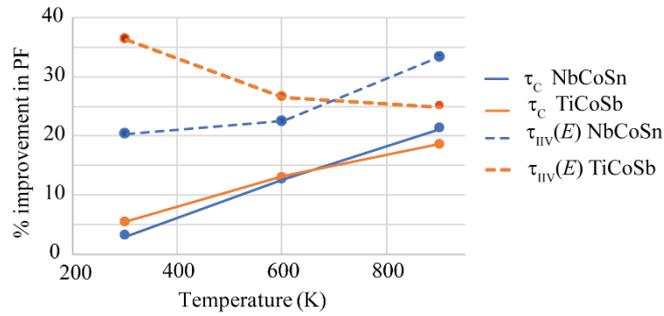

**Fig. 8. Percentage improvement of the power factor (from peak to peak between unstrained and 5% compression) under a constant rate of scattering (solid-lines), and energy dependent $\tau_{IIV}$ scattering(dashed-lines) versus temperature for NbCoSn, and TiCoSb.**

# Bibliography

1.      Beretta, D. *et al.* Thermoelectrics: From history, a window to the future. *Mater. Sci. Eng. R Reports* (2018). doi:10.1016/J.MSER.2018.09.001

2.      Bell, L. E. Cooling, heating, generating power, and recovering waste heat with thermoelectric systems. *Science* **321,** 1457–61 (2008).

3.      Shakouri, A. Recent Developments in Semiconductor Thermoelectric Physics and Materials. *Annu. Rev. Mater. Res.* **41,** 399–431 (2011).

4.      Koumoto, K. *et al.* Thermoelectric Ceramics for Energy Harvesting. *J. Am. Ceram. Soc.* **96,** 1–23 (2013).

5.      Norouzzadeh, P. & Vashaee, D. Classification of Valleytronics in Thermoelectricity. *Sci. Rep.* **6,** 22724 (2016).

6.      Pei, Y. *et al.* Convergence of electronic bands for high performance bulk thermoelectrics. *Nature* **473,** 66–69 (2011).

7.      Bhattacharya, S. & Madsen, G. K. H. High-throughput exploration of alloying as design strategy for thermoelectrics. *Phys. Rev. B* **92,** 85205 (2015).

8.      Tan, G., Zhao, L.-D. & Kanatzidis, M. G. Rationally Designing High-Performance Bulk Thermoelectric Materials. *Chem. Rev.* **116,** 12123–12149 (2016).

9.      Lee, J.-H. Significant enhancement in the thermoelectric performance of strained nanoporous Si. *Phys. Chem. Chem. Phys.* **16,** 2425–2429 (2014).

10.     Huang, L. *et al.* Recent progress in half-Heusler thermoelectric materials. *Mater. Res. Bull.* **76,** 107–112 (2016).

11.     Page, A., Poudeu, P. F. P. & Uher, C. A first-principles approach to half-Heusler thermoelectrics: Accelerated prediction and understanding of material properties. *J. Mater.* **2**, 104–113 (2016).

12.     Xie, W. *et al.* Recent Advances in Nanostructured Thermoelectric Half-Heusler Compounds. *Nanomaterials* **2**, 379–412 (2012).

13.     Bos, J.-W. G. & Downie, R. A. Half-Heusler thermoelectrics: a complex class of materials. *J. Phys. Condens. Matter* **26**, 433201 (2014).

14.     Casper, F., Graf, T., Chadov, S., Balke, B. & Felser, C. Half-Heusler compounds: novel materials for energy and spintronic applications. *Semicond. Sci. Technol.* **27**, 063001 (2012).

15.     Scheidemantel, T. J., Ambrosch-Draxl, C., Thonhauser, T., Badding, J. V & Sofo, J. O. Transport coefficients from first-principles calculations. *Phys. Rev. B* **68**, 125210 (2003).

16.     Yang, J. *et al.* Evaluation of Half-Heusler Compounds as Thermoelectric Materials Based on the Calculated Electrical Transport Properties. *Adv. Funct. Mater.* **18**, 2880–2888 (2008).

17.     Kumarasinghe, C. & Neophytou, N. Band alignment and scattering considerations for enhancing the thermoelectric power factor of complex materials: The case of Co-based half-Heusler alloys. *Phys. Rev. B* **99**, 195202 (2019).

18.     Mahan, G. D., Balseiro, Ti. & Atomico Bariloche, C. The Best Thermoelectric. *Proc. Natl. Acad. Sci.* **93**, 7436–7439 (1996).

19.     Neophytou, N., Wagner, M., Kosina, H. & Selberherr, S. Analysis of Thermoelectric Properties of Scaled Silicon Nanowires Using an Atomistic Tight-Binding Model. *J. Electron. Mater.* **39**, 1902–1908 (2010).

20. Giannozzi, P. *et al.* QUANTUM ESPRESSO: a modular and open-source software project for quantum simulations of materials. *J. Phys. Condens. Matter* **21,** 395502 (2009).

21. Corps, J. *et al.* Interplay of Metal-Atom Ordering, Fermi Level Tuning, and Thermoelectric Properties in Cobalt Shandites $Co_3M_2S_2$ (M = Sn, In). *Chem. Mater.* **27,** 3946–3956 (2015).

22. Roychowdhury, S., Shenoy, U. S., Waghmare, U. V. & Biswas, K. An enhanced Seebeck coefficient and high thermoelectric performance in p-type In and Mg co-doped $Sn_{1-x}Pb_xTe$ via the co-adjuvant effect of the resonance level and heavy hole valence band. *J. Mater. Chem. C* **5,** 5737–5748 (2017).

23. Capellini, G. *et al.* Tensile Ge microstructures for lasing fabricated by means of a silicon complementary metaloxide-semiconductor process. *Opt. Express* **22,** 399–410 (2014).

24. Jeong, I., Kwon, J., Kim, C. & Park, Y. J. Design and numerical analysis of surface plasmon-enhanced fin Ge-Si light-emitting diode. *Opt. Express* **22,** 5927 (2014).

25. Low, K. L., Yang, Y., Han, G., Fan, W. & Yeo, Y.-C. Electronic band structure and effective mass parameters of Ge 1-x Sn x alloys. *Cit. J. Appl. Phys.* **112,** 73707 (2012).

26. Zhou, J., Cheng, S., You, W.-L. & Jiang, H. Effects of intervalley scattering on the transport properties in one−dimensional valleytronic devices. *Sci. Rep.* **6,** 23211 (2016).